\documentclass[sigconf]{acmart}

\usepackage{multirow, makecell}
\usepackage{multicol, makecell}
\usepackage{mathtools, nccmath}

\usepackage{mathtools}
\DeclarePairedDelimiter\ceil{\lceil}{\rceil}

\AtBeginDocument{%
  \providecommand\BibTeX{{%
    \normalfont B\kern-0.5em{\scshape i\kern-0.25em b}\kern-0.8em\TeX}}}

\copyrightyear{2021}
\acmYear{2021}
\setcopyright{acmcopyright}\acmConference[MICRO '21]{MICRO-54: 54th Annual IEEE/ACM International Symposium on Microarchitecture}{October 18--22, 2021}{Virtual Event, Greece}
\acmBooktitle{MICRO-54: 54th Annual IEEE/ACM International Symposium on Microarchitecture (MICRO '21), October 18--22, 2021, Virtual Event, Greece}
\acmPrice{15.00}
\acmDOI{10.1145/3466752.3480052}
\acmISBN{978-1-4503-8557-2/21/10}

\begin{document}

\title{HiMA: A Fast and Scalable History-based Memory Access Engine for Differentiable Neural Computer}

\author{Yaoyu Tao, Zhengya Zhang}
\email{{taoyaoyu, zhengya}@umich.edu}
\affiliation{%
  \institution{University of Michigan}
  \city{Ann Arbor}
  \state{Michigan}
  \country{USA}
  \postcode{48109}
}

\begin{abstract}
Memory-augmented neural networks (MANNs) provide better inference performance in many tasks with the help of an external memory. The recently developed differentiable neural computer (DNC) is a MANN that has been shown to outperform in representing complicated data structures and learning long-term dependencies. DNC's higher performance is derived from new history-based attention mechanisms in addition to the previously used content-based attention mechanisms. History-based mechanisms require a variety of new compute primitives and state memories, which are not supported by existing neural network (NN) or MANN accelerators. We present HiMA, a tiled, history-based memory access engine with distributed memories in tiles. HiMA incorporates a multi-mode network-on-chip (NoC) to reduce the communication latency and improve scalability. An optimal submatrix-wise memory partition strategy is applied to reduce the amount of NoC traffic; and a two-stage usage sort method leverages distributed tiles to improve computation speed. To make HiMA fundamentally scalable, we create a distributed version of DNC called DNC-D to allow almost all memory operations to be applied to local memories with trainable weighted summation to produce the global memory output. Two approximation techniques, usage skimming and softmax approximation, are proposed to further enhance hardware efficiency. HiMA prototypes are created in RTL and synthesized in a 40nm technology. By simulations, HiMA running DNC and DNC-D demonstrates 6.47$\times$ and 39.1$\times$ higher speed, 22.8$\times$ and 164.3$\times$ better area efficiency, and 6.1$\times$ and 61.2$\times$ better energy efficiency over the state-of-the-art MANN accelerator. Compared to an Nvidia 3080Ti GPU, HiMA demonstrates speedup by up to 437$\times$ and 2,646$\times$ when running DNC and DNC-D, respectively.
\end{abstract}

\begin{CCSXML}
<ccs2012>
<concept>
<concept_id>10010520.10010521.10010542.10010294</concept_id>
<concept_desc>Computer systems organization~Neural networks</concept_desc>
<concept_significance>500</concept_significance>
</concept>
</ccs2012>
\end{CCSXML}

\ccsdesc[500]{Computer systems organization~Neural networks}
\ccsdesc[500]{Computer systems organization~Data flow architectures}
\ccsdesc[500]{Computer systems organization~Special purpose systems}

\keywords{differentiable neural computer, memory-augmented neural networks, memory access engine}

\maketitle

\section{Introduction}

The application of neural networks (NNs) have grown extensively to many practical problems such as natural language processing (NLP) \cite{nn_nlp}, speech recognition \cite{nn_speech} and computer vision (CV) \cite{nn_cv}. In the case of NLP, the improvements come from the sequence modeling capabilities of recurrent NNs (RNNs) \cite{nlp_rnn} such as long short-term memory (LSTM) \cite{nlp_lstm} or gated recurrent units (GRU) \cite{gru_paper}. However, the performance of RNNs is limited by how long memories can persist, because the dynamic states are intrinsically embodied within the network. RNNs become less effective in tasks like question answering (QA) \cite{mem_net_qa} where relevant information for the correct answers could be far away from where the questions are asked. This motivated the development of memory-augmented NNs (MANNs), a fully differentiable model that contains an isolated external memory module that NNs can learn to store to and read from when computing predicted outputs. 

Such MANNs include memory network (MemNet and MemN2N) \cite{mem_net_qa, mem_net}, dynamic memory network (DMN) \cite{dmn_paper}, neural Turing machine (NTM) \cite{ntm_paper} and differentiable neural computer (DNC) \cite{dnc_paper}. Compared to traditional RNN/LSTM or recently developed Transformer \cite{vaswani2017attention}, MANNs outperform in tackling long-term dependency problems and find many applications not only in NLP, but also in graph modeling \cite{oh2016control,antoniou2017data}, navigation \cite{dnc_paper, singla2019memory} and reinforcement learning \cite{borsa2017observational,foerster2017stabilising,clavera2018model}. Specifically, DMN uses a GRU as the memory component. MemNet/MemN2N uses an external addressable memory; however, neither memory content nor access history is considered in the addressing. NTM enhances the performance by using content-based soft write and read. An NTM can infer simple algorithms such as copying or sorting. Subsequently, DNC extends NTM by incorporating history-based attention mechanisms that consider historical events when accessing external memory. This allows DNC to achieve better performance than NTM in handling long-term dependencies \cite{dnc_paper}. However, the enhanced performance of DNC comes at a high computational cost, complex memory operations, and specifically history-based attention mechanisms.

NN accelerators \cite{fused_cnn_micro_2016, dnn_dataflow_micro_2019,dnn_micro_2018,scnn_2017,simba,kwon2019herald_heteroDNN} cannot run DNC due to the lack of capability to handle elaborate memory operations. Compared to NN accelerators that store weights, perform convolutions and accumulate partial sums, DNC accelerators need to support more complex and diverse workloads, including new primitives like sorting and matrix transpose and a variety of new state memories to store access history that do not exist in NNs. MANN accelerators for MemNets or NTM \cite{mnnfast,mnn_accelerator_paper,xmann,manna} also cannot run DNC due to the lack of support of DNC's unique sorting primitive and new state memories. The only way to run DNC using an existing accelerator is to have it attached to a general-purpose CPU or GPU, which is unlikely to deliver a high efficiency.

NTM accelerators only support content-based attention mechanisms.
%without state memories.
Specifically, X-MANN \cite{xmann} implements external memory using resistive crossbars. The performance gain relies on emerging devices that are not widely available. The recently developed \textsc{M\scriptsize ANNA} \cite{manna} proposed a network-on-chip (NoC) architecture for NTM. \textsc{M\scriptsize ANNA}'s distributed architecture provides more memory bandwidth and compute parallelism, but its H-tree NoC still incurs a traffic bottleneck when running DNC's history-based attention mechanisms. DNC accelerators have been developed recently \cite{dnc_fpga,dnc_farm}. In \cite{dnc_farm}, the efficiency mainly comes from analog-based processing elements such as analog-to-digital converters (ADCs), which are more sensitive to variations and noise, and less portable between process technologies. These designs all followed a centralized architecture for memory access and compute, which could lead to poor scalability when memory size increases. Operations may need to be serialized, degrading both speed and latency.

We present HiMA, a \underline{Hi}story-based \underline{M}emory \underline{A}ccess engine to efficiently accelerate DNC memory operations. To the best of our knowledge, HiMA is the first distributed, tiled architecture that supports all DNC features. History-based attention mechanisms introduce a variety of new primitives and state memories, requiring access to various memories concurrently and incurring complex traffic in an NoC architecture. HiMA focuses on distributed processing of DNC and optimizing NoC traffic to enhance scalability. We summarize the contributions of this work as follows:

\begin{itemize} 

\item \emph{Scalable Multi-Mode NoC}. We study the DNC computation and memory access profile, especially history-based attention mechanisms. Based on the analysis, a multi-mode NoC is designed to adapt to DNC's traffic profile, improving both traffic latency and scalability.

\item \emph{Optimized Memory Partition}. Conventional row or column-wise partition is suboptimal for DNC's new state memories. We consider both content-based and history-based mechanisms and propose a submatrix-wise partition to reduce NoC traffic.

\item \emph{Distributed and Efficiency-Enhanced Kernels}. Both memory and memory operations are distributed to tiles using a new distributed DNC (DNC-D) model. The distributed computational kernels minimize the NoC traffic and provide a higher parallelism. We propose local-global two-stage sort, usage skimming, and softmax approximation to reduce the complexity and improve the computational efficiency.

\end{itemize}

Prototypes of HiMA are implemented in RTL and synthesized in a 40nm technology. HiMA running DNC and DNC-D demonstrates up to 6.47$\times$ and 39.1$\times$ improvements in speed, 22.8$\times$ and 164.3$\times$ improvements in area efficiency, and 6.1$\times$ and 61.2$\times$ improvements in energy efficiency, respectively, over \textsc{M\scriptsize ANNA}, the state-of-the-art MANN accelerator for NTM. Compared to an Nvidia 3080Ti GPU, HiMA shortens the inference time by up to 437$\times$ and 2,646$\times$ when running DNC and DNC-D.

\section{Background}

%We provide a brief overview of relevant concepts, mathematical descriptions of DNC and review  MANN accelerators. 

%\subsection{Memory-Augmented Neural Networks}

\begin{figure}
\centering
\includegraphics[width=0.90\linewidth]{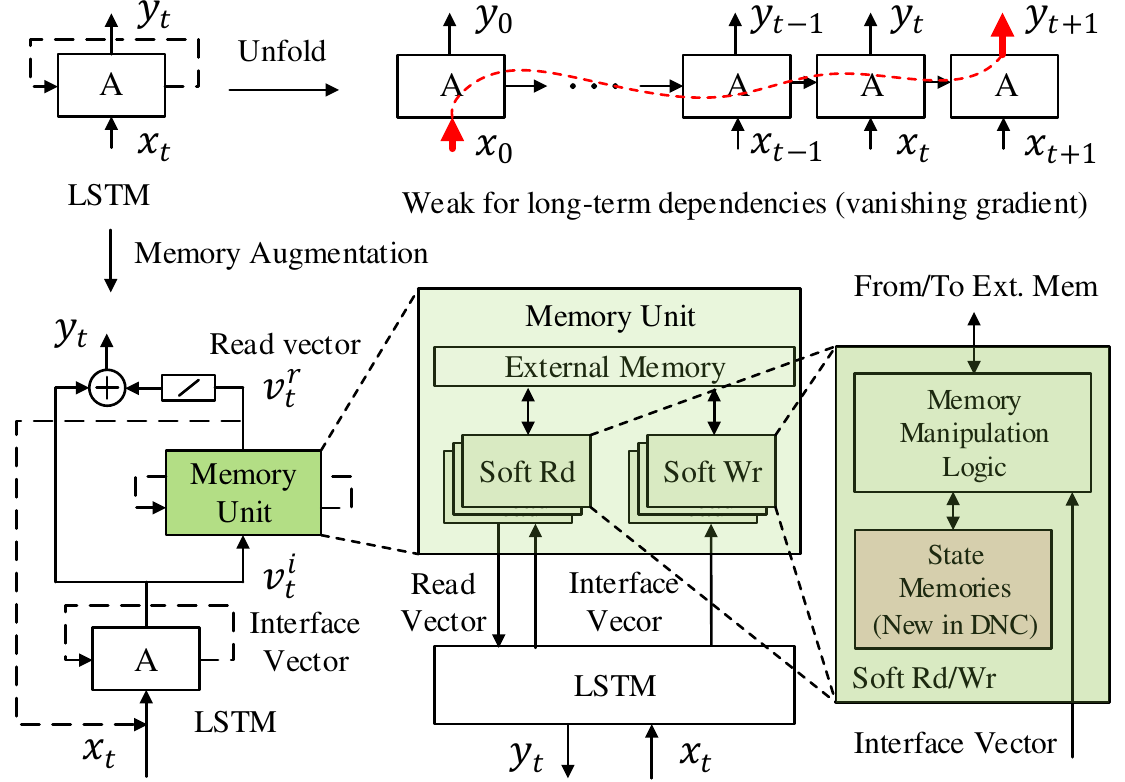}
\caption{LSTM and memory-augmented neural network.}
\label{mann}
\end{figure}

Recurrent neural networks (RNNs) \cite{rnn_first_paper} such as LSTMs \cite{lstm_first_paper} extend feedforward DNNs by introducing recurrent connections, thereby allowing the networks to store dynamic states across iterations of inputs. Let $x_t$ be the input and $y_t$ be the output at time $t$, an LSTM is composed of a chain of LSTM cells, denoted by $A$, as shown in Figure~\ref{mann}. The introduction of dynamic state has benefited domains which require remembering of event sequences, such as QA in NLP. However, the amount of information that can be stored in the network is bounded by the size of the underlying network. Therefore LSTM lacks the scalability to handle complicated sequence events with long-term dependencies.

\begin{figure*}
\centering
\includegraphics[width=\linewidth]{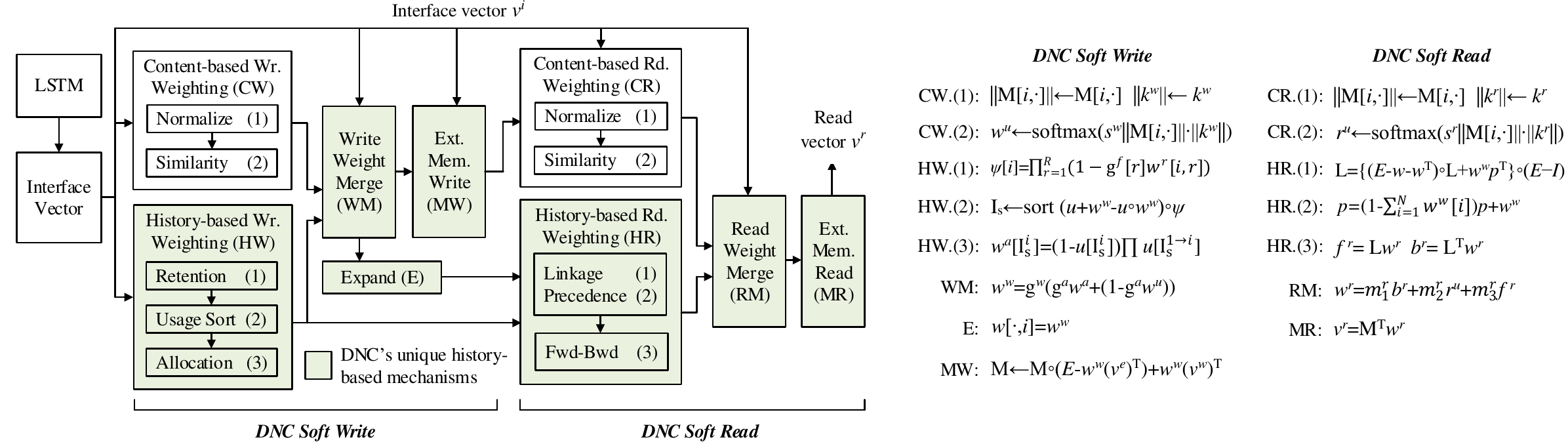}
\caption{DNC inference dataflow: history-based and content-based memory access in DNC memory unit.}
\label{dnc}
\end{figure*}

To address the scaling problem of LSTM, MANNs have been proposed as shown in Figure~\ref{mann}. A memory unit is connected to an NN (typically an LSTM) and the external memory\footnote{In the context of MANN, an external memory refers to a memory external to the NN (e.g., LSTM) that stores data.} can be accessed by attention-based mechanisms through soft read and soft write heads. Specifically, at time $t$ the LSTM sends an \textit{interface vector} $v_t^i$ to the memory unit and receives a \textit{read vector} $v_t^r$ from the memory unit. In this way, the dynamic state can be explicitly decoupled from the neural network. NTM \cite{ntm_paper} is a MANN that outperforms LSTM by employing content-based soft write and read to access the memory, a form of attention mechanism. However, memory access history is completely discarded in memory slot selection and weighting. DNC \cite{dnc_paper} extends NTM and the memory is accessed based on both memory content and memory access history. History-based attention mechanisms allow DNC to achieve much better performance than NTM, but also introduce complex memory manipulations and a variety of new state memories as highlighted in Figure~\ref{mann}.

\subsection{Differentiable Neural Computer}

In this work, we focus on DNC memory unit where elaborate memory operations take place. The DNC inference dataflow and mathematical descriptions of the operations are shown in Figure~{\ref{dnc}}. Compared to other variants of MANNs such as NTM or MemNet, DNC is the only model that incorporates history-based memory access. The NN (e.g., LSTM) sends an input to the memory unit, known as the \emph{interface vector} $v^i$, including the necessary access information such as write key $k^w$, read key $k^r$ or write vector $v^w$. The memory unit returns the \emph{read vector} $v^r$. Suppose the memory $M$ is modeled as a $N\times W$ matrix ($N > W$) and the number of read heads (i.e., number of parallel reads) is $R$, we provide a brief operational explanation of the DNC soft write and soft read.

\subsubsection{Soft Write}

As illustrated in Figure~\ref{dnc}, soft write is done in two steps: 1) compute write weighting $w^w$, and 2) memory write, i.e., apply the weighting $w^w$ to the values (write vector $v^w$ and erase vector $v^e$) and write them to memory.

In DNC, the write weighting is a combination of \emph{content-based weighting} and \emph{history-based weighting}. The content-based weighting is inherited from NTM, and it is based on the similarity to the write key. Mathematically, the memory entries $M[i,\cdot]$ and the write key $k^w$ are first normalized, and the similarity between the two is computed as the content-based write weighting $w^u$.

The history-based weighting is brand new in DNC. DNC enhances the selection of memory cells by biasing towards those that are most recently read from (based on read weighting $w^r$ from the previous time step), least recently written to (based on write weighting $w^w$ from the previous time step), or deemed inconsequential (based on free gate $g^f$). The history-based weighting is computed in three steps: 1) the retention vector $\psi$ is first calculated based on the free gate $g^f$ and the read weighting $w^r$; 2) the \emph{usage vector} $u$ is updated based on the retention vector and the write weighting $w^w$, and then sorted; and 3) the history-based write weighting $w^a$ is computed by accumulating the product of the sorted usage. The content-based weighting $w^u$ and the history-based weighting $w^a$ are combined to obtain the write weighting $w^w$.

%The write weighting is applied to the write vector $v^w$ and the erase vector $v^e$ before the vectors are written to memory.
%where $E$ is an $N\times N$ matrix of ones.

\subsubsection{Soft Read}

Soft read is done in two steps: 1) compute read weighting $w^r$, and 2) memory read, i.e., apply the weighting $w^r$ to memory $M$ to obtain the read vector $v^r$.

Similar to soft write, soft read combines both content-based weighting and history-based weighting. The content-based read weighting $r^u$ is computed in the same way as the content-based write weighting.
The history-based read weighting is computed in three steps: 1) the write weighting $w^w$ is first expanded to an $N\times N$ matrix to derive linkage matrix $L$. The \emph{linkage} tracks the order in which memory locations are written to;
%The $E-I$ term in its calculation removes self-links where $I$ is the identity matrix.
2) the precedence vector $p$ is updated to track the degree each memory entry is most recently written to; and 2) a \emph{forward and backward pass} is used to merge the read weighting $w^r$ from the previous time step with the linkage matrix $L$, as well as the content-based read weighting $r^u$ to update the read weighting $w^r$.

\begin{table*}
	\centering
	\caption{Analysis of DNC Kernels}
	\renewcommand{\arraystretch}{1.1}
	\begin{tabular}{|c|c|c|c|c|c|c|c|}
        \hline
		\multirowcell{2}{Type} & \multirowcell{2}{Category} & \multirowcell{2}{Kernel Name} & \multirowcell{2}{Key Primitives} & Ext. Mem & State Mem & Total NoC \\ 
		& & & & Access & Access & Traffic \\ \hline \hline
		\multirowcell{4}{Access\\Kernels} & \multirowcell{2}{Content-based\\Weighting} & Normalize & inner-prod & $O(NW)$ & $O(W)$ & $O(N_tN)$ \\ \cline{3-7}
	    & & Similarity & inner-prod & $O(NW)$ & $O(W)$ & $O(N_t)$ \\ \cline{2-7}
	    & \multirowcell{2}{Memory\\Access} &  Memory Write & el-add/sub/mult, outer-prod & $O(NW)$ & $O(N)$ & $O(N_tN)$ \\ \cline{3-7}
	    & & Memory Read & transpose, mat-vec mult & $O(NW)$ & $O(N)$ & $O(N_tNW)$ \\ \hline \hline
	    \multirowcell{9}{State\\Kernels\\ (\textit {New in} \\ \textit {DNC})} &  \multirowcell{5}{History-based\\Write\\Weighting} & Retention & el-mult, vec acc-prod & No & $O(RN)$ & No\\ \cline{3-7}
	    & & Usage & el-add/sub/mult & No & $O(N)$ &  No\\ \cline{3-7}
	    & & Usage Sort & sort (Section~\ref{compute_optimization}) & No & $O(N)$ &  $O(N)$ \\ \cline{3-7}
	    & & Allocation & vec acc-prod & No & $O(N)$ & $O(N_t)$\\ \cline{3-7}
	    &  & Wr. Weight Merge & el-add/sub & No & $O(N)$ & No \\ \cline{2-7}
	    & \multirowcell{5}{History-based\\Read\\Weighting} & \multirowcell{2}{Linkage} & mat expand, outer-prod, & \multirowcell{2}{No} & \multirowcell{2}{$O(N^2)$} & \multirowcell{2}{$O(N_tN)$} \\
	    & & & el-add/sub/mult & & & \\ \cline{3-7}
	    & & Precedence & el-add, vec acc-sum & No & $O(N)$ & $O(N_t)$  \\ \cline{3-7}
	    & & Forward-backward & transpose, mat-vec mult & No & $O(N^2)$ & $O(N_tN^2)$ \\ \cline{3-7}
	    &  & Rd. Weight Merge & el-add & No & $O(RN)$ & No\\ \hline
	\end{tabular}
	\label{tbl:dnc_kernel_analysis}
\end{table*}

\subsection{State-of-the-art MANN Accelerators}

\begin{figure}
\centering
\includegraphics[width=0.98\linewidth]{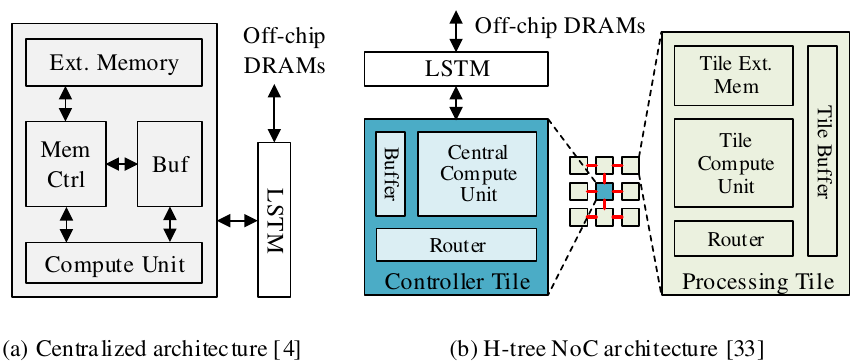}
\caption{Centralized-memory architecture \cite{dnc_farm} and tiled architecture \cite{manna} for accelerating MANN's memory unit.}
\label{mann_acc}
\end{figure}

Conventional NN or matrix multiplication accelerators do not fully support DNC because they lack the primitives including sorting and matrix transpose, and miss a specialized memory unit to support a variety of state memories. MANN accelerators \cite{mnn_accelerator_paper,mnnfast,manna,xmann,dnc_farm,dnc_fpga} have been proposed for MANN's memory unit. The memory unit receives an input interface vector from an NN accelerator that executes LSTM inference. In performing the inference, the NN accelerator may communicate with off-chip DRAMs.

Here we focus on NTM and DNC accelerators that support both soft write and soft read and omit simpler accelerators that do not come with such support. Some NTM and DNC accelerators use a centralized-memory architecture \cite{xmann,dnc_farm,dnc_fpga} as shown in Figure~\ref{mann_acc}(a). In particular, \cite{xmann,dnc_farm} improve the memory access efficiency by introducing in-memory compute through resistive crossbar or analog operations. However, the centralized memory ultimately limits the bandwidth and parallelism available, and the emerging devices and custom mixed-signal circuits are not yet practical for sufficiently large memory sizes.

\textsc{M\scriptsize ANNA} \cite{manna} introduces the first tiled NoC architecture as shown in Figure~\ref{mann_acc}(b) to solve the bandwidth and parallelism limitation. \textsc{M\scriptsize ANNA} contains two types of tiles: processing tile (PT) which includes external memory sub-banks and the associated compute units, and controller tile (CT) which includes top-level processing units to distribute information to the PTs and collect results from the PTs. Designed for NTM, \textsc{M\scriptsize ANNA} cannot run DNC due to the lack of support for new primitives like sorting and new state memories for maintaining access history. \textsc{M\scriptsize ANNA}'s H-tree NoC also becomes less efficient in carrying inter-tile communication when the PT count increases. The inefficiency is exacerbated by DNC's history-based attention mechanisms that inject complex traffic patterns onto the NoC, limiting its scalability as seen in Figure~\ref{noc_topology_arch}(d).

\section{Analysis of DNC Kernels}
\label{dnc_acc_challenges_section}

We first analyze DNC's memory access and computational profile, followed by a simulation study of DNC running bAbI dataset \cite{babi_dataset} on CPU and GPU. 

\subsection{Theoretical Kernel Analysis}

Table~\ref{tbl:dnc_kernel_analysis} lists DNC's computational kernels with their corresponding primitives, associated memory access complexity and the NoC traffic condition when mapped to a tiled architecture. Recall that the external memory $M$ is modeled as a $N\times W$ matrix ($N > W$) and the number of read heads is $R$. Let $N_t$ be the number of tiles in a tiled architecture. We categorize DNC kernels into two types: 1) state kernels for maintaining memory states and determining how the external memory is accessed, and 2) access kernels that do not maintain states and perform the actual access to the external memory. NTM only needs access kernels, while DNC requires a variety of new state kernels to support history-based mechanisms. These new state kernels impose critical challenges:

\begin{itemize}
    \item \textit{Computation}: Kernels such as usage sort and forward-backward require compute-intense large-scale data sorting of $O(N\text{log}N)$ complexity (assume merge sort) or matrix-vector multiplication of $O(N^2)$ complexity, which can be major bottlenecks.
    
    \item \textit{Memory Access}: Kernels such as linkage and forward-backward require $O(N^2)$ accesses to the new linkage memory, which easily surpass the memory access by the access kernels used by other MANNs like NTM. 
    
    \item \textit{NoC Traffic}: In a tiled architecture, some kernels such as linkage rely on inter-tile traffic. The amount of NoC traffic can be as high as $O(N_tN^2)$ depending on state and external memory partitions, and the traffic pattern is non-uniform over time and space due to the different primitives and state memories involved. 
\end{itemize}

The challenges require designing an efficient NoC and memory organization, maximizing distributed processing, and providing efficient computational kernels such as sort.

\subsection{Kernel Runtime Analysis}

\begin{figure}
\centering
\includegraphics[width=0.98\linewidth]{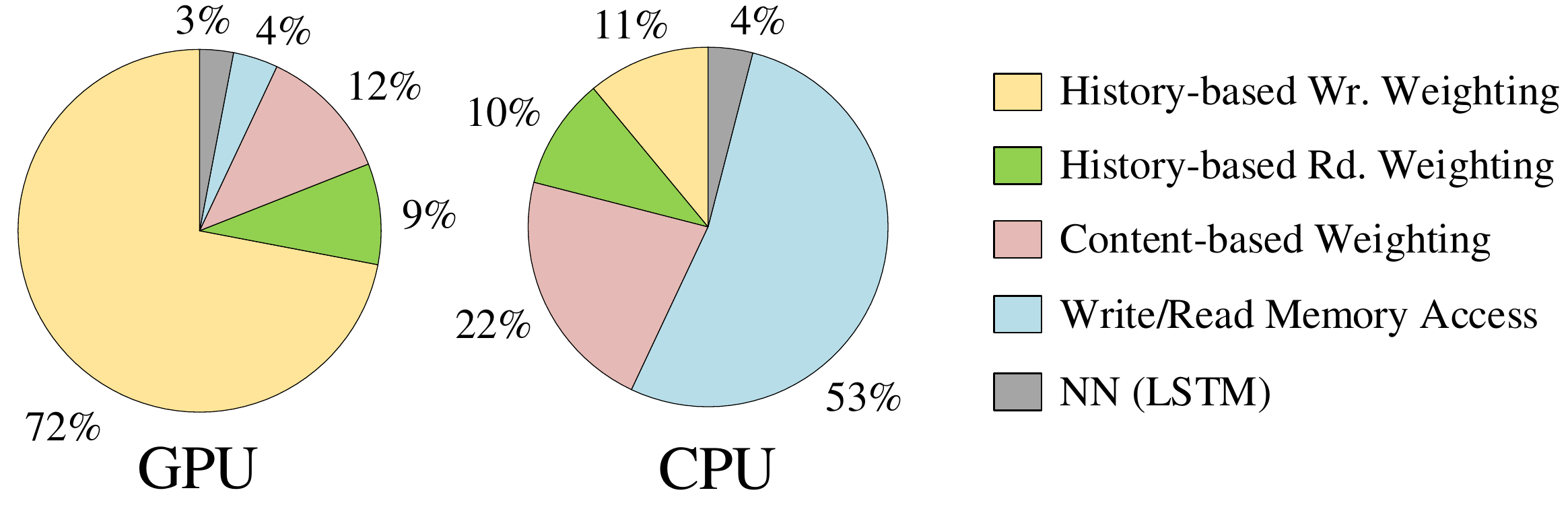}
\caption{Kernel runtime breakdown on CPU/GPU for the bAbI dataset. The external memory size is $N\times W = 1024\times64$ and the LSTM is 1-layer of size 256.} 
\label{kernel_breakdown}
\end{figure}

We simulated DNC inference on an Nvidia 3080Ti GPU and an Intel Core i7-9700K CPU using the bAbI dataset \cite{babi_dataset} in NLP. The bAbI dataset consists of 20 tasks. Each task is independent of the others, tests one aspect of an intended NLP behavior and includes as many as 10,000 QA examples. It is the only publicly available and practically meaningful dataset to date to demonstrate DNC's performance. In our experiments, we ran tasks in the bAbI dataset and recorded the average runtime. Figure~\ref{kernel_breakdown} captures runtime breakdown of DNC kernels in different categories. The average GPU inference time is 5.16~ms/test, 2.12$\times$ faster than the 10.94~ms/test inference time on the CPU. On both the GPU and the CPU, the NN (an LSTM) takes less than 5\% of the total runtime, while the memory unit takes more than 95\% of the total runtime. \textit{It highlights the need of a memory access engine for DNC.} Additional insights can be derived from the kernel runtime shown in Figure~\ref{kernel_breakdown}.

\begin{itemize}
    \item History-based write weighting, including retention, usage sort and allocation, accounts for 72\% of the runtime on the GPU. A possible explanation is that GPU is not well suited for speeding up large-scale sorting. %efficiently.
    %and the performance degrades further for a larger memory.
    \item History-based read weighting, including linkage and precedence, relies on vector and matrix operations that can be extensively parallelized by the GPU. This part uses only 9\% of the GPU's runtime.
    \item Content-based write/read weighting, including normalization and similarity, involves many multiply-accumulate (MAC) and softmax operations. It costs 12\% of the runtime on the GPU and 22\% on the CPU, but it is not the dominant part on either platform.
    \item Memory write and read\footnote{In the context of DNC, memory write and memory read are not simply write to or read from memory. Applying write weighting to data before write to memory and applying read weighting to data read from memory are the dominant operations in memory write and memory read, respectively.} are much faster on the GPU (4\% of the runtime) than on the CPU (53\% of the runtime), because they are dominated by parallelizable weighting operations using MAC arrays.
\end{itemize}

The results show that improving the DNC performance requires both optimized computational kernels such as sort and highly parallel matrix operations. Another consideration is that GPU and CPU follow a centralized-memory architecture and high premiums in power and area are paid in sustaining a high-bandwidth interface and a versatile memory hierarchy. In designing an accelerator targeting high performance, high energy efficiency and low cost, a distributed, tiled architecture is the preferred approach.

\section{HiMA Architecture Design}

HiMA is a memory access engine that follows a distributed, tiled architecture that consists of one CT and many PTs with an NoC linking the tiles. Based on kernel analysis and simulation results, we summarize HiMA's architectural design goals:

\begin{itemize}
    \item \emph{Scalable and versatile NoC}: \label{scale_section}
    A fixed NoC is sub-optimal for the complex NoC traffic by DNC's unique state kernels. The H-tree NoC in \cite{manna} suffers from traffic saturation with more than 8 tiles as shown in {Figure~\ref{noc_topology_arch}(d)}. A more scalable NoC that supports DNC's versatile kernels is desired.

    \item \emph{Efficient memory partition}: DNC's external memory is large, and its state memories can be even larger. For example, the linkage matrix requires a memory of $N\times N$ ($N > W$).
    A strategy is needed to partition large external and state memories and distribute them to tiles with the goal of minimizing the NoC traffic amount when executing DNC kernels.
    %\textsc{M\scriptsize ANNA} \cite{manna} applies a simple row-wise partition to distribute external memory while state memories are not supported. We revisit memory partition, especially for state memories, to propose a new approach to minimize the amount of NoC traffic.

    \item \emph{Distributed compute kernels}: DNC's kernels are ideally distributed to the tiles. The sort kernel is especially important. It is a performance bottleneck and it is not supported by existing NN/MANN accelerators \cite{fused_cnn_micro_2016, dnn_dataflow_micro_2019,dnn_micro_2018,scnn_2017,simba, kwon2019herald_heteroDNN, manna,xmann,mnnfast,mnn_accelerator_paper}.
    The goal is to distribute such kernels along with memory to tiles, while minimizing the NoC traffic amount.
    %A conventional centralized sorting requires all PTs to send the local vectors to the CT. However, it is inefficient and slow as the CT becomes the performance bottleneck with increasing PT count. Designing a sort kernel needs to consider how to make it distributed.
\end{itemize}

\subsection{Scalable Multi-Mode NoC}
\label{noc_optimization}

\begin{figure}
\centering
\includegraphics[width=0.95\linewidth]{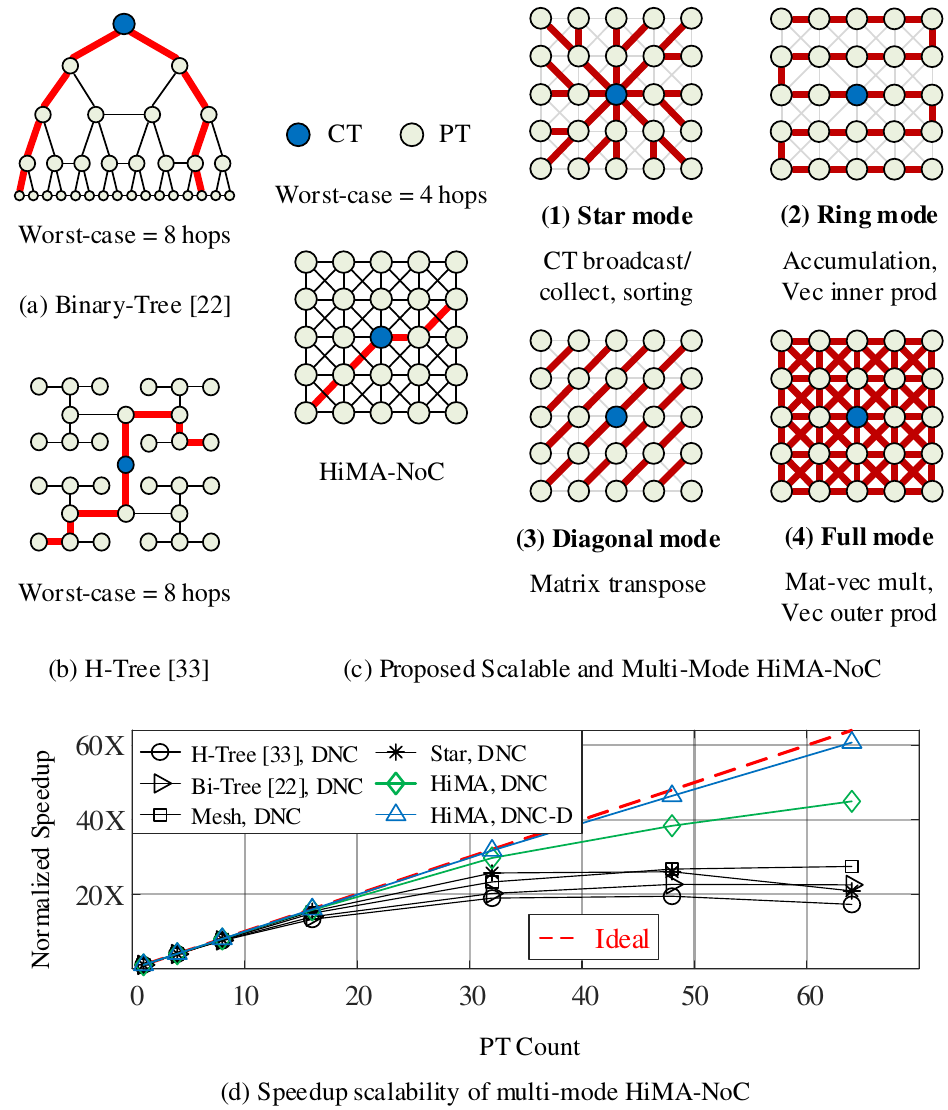}
\caption{Scalable and Multi-Mode HiMA-NoC.}
\label{noc_topology_arch}
\end{figure}

Reconfigurable NoCs have been proposed for DNN accelerators for different tensor sizes. Specifically, \textsc{M\scriptsize AERI} \cite{kwon2018maeri_reconf_interconnect} and \textsc{H\scriptsize ERALD} \cite{kwon2019herald_heteroDNN} employ a multi-layer binary tree as show in Figure~\ref{noc_topology_arch}(a) with configurable interconnects between adjacent sub-trees at each level. These designs are suitable for DNN's reduction, collection and multi-cast dataflows. They are however sub-optimal for DNC's diverse traffic patterns, especially for transpose and matrix-vector multiplication that require communications between distant tiles. Data transfer between two distant tiles may need to pass through their mutual root tile, which can become a traffic congestion point that drastically increases the inter-tile communication latency. The H-tree NoC demonstrated by \textsc{M\scriptsize ANNA} \cite{manna} was designed for access kernels that require only two types of inter-tile communication: 1) broadcast of interface vectors from CT to PTs, and collection of read vectors from the PTs; and 2) transfer of submatrices of the external memory and partial sums between PTs for transpose and matrix-vector multiplication. In implementing the access kernels, the H-tree NoC does not present a bottleneck up to 16 tiles \cite{manna}.

For history-based memory access, the new state kernels introduce far more inter-tile traffic and a diverse traffic profile. To see the suitability of the H-tree NoC and the multi-layer binary tree, we mapped DNC to a tiled architecture utilizing these two NoCs. To check scalability, we simulated the speedup by increasing the number of PTs. Here we assume ideal CT and PTs where the memory bandwidth or the computational parallelism do not present bottlenecks, and ideal routers that can handle any traffic congestion by stalling. Figure~\ref{noc_topology_arch}(d) shows that the speedup starts to saturate beyond the 8 tiles for both NoCs. The H-tree NoC \cite{manna} requires traffic between two tiles to go through their mutual root tile, as show in Figure~\ref{noc_topology_arch}(b), resulting in traffic congestion at highest root node for the distant pairs of PTs. The binary-tree NoC \cite{kwon2018maeri_reconf_interconnect} is an enhanced H-tree by additional interconnects between adjacent sub-trees at each level, as shown in Figure~\ref{noc_topology_arch}(a). It outperforms the H-tree slightly, but its scalability still saturates at a low level.
We analyze the traffic profile of DNC primitives and the suitable NoC topologies.

\begin{itemize}
    \item Interface vector broadcast, read vector collection and sorting require only CT-PT traffic. A star NoC is the most suitable, where all PTs are connected directly to the CT with a distance of 1 hop. However, the CT needs a complex router and can become a traffic congestion point with increasing PT count, limiting scalability.
    \item Accumulation of products or sums and vector inner product require sending accumulated results from one PT to the next PT. A ring NoC is the most suitable.
    \item Matrix transpose requires transferring on-tile submatrices to other tiles along diagonals, as shown in Figure~\ref{noc_topology_arch}(c). A diagonally-connected NoC is suitable.
    \item Matrix vector multiplication and vector outer product require each tile to send its local submatrices to all other tiles for computation. A full-duplex mesh NoC is the most suitable. However, the scalability of a full-duplex mesh is even worse than the star NoC, because every tile, not only CT, can become a traffic congestion point.
\end{itemize}

The analysis shows that a fixed NoC topology does not meet DNC's diverse traffic profile. We propose a multi-mode NoC, named HiMA-NoC, to shorten the transfer distance, reduce the traffic congestion and enhance the scalability. Figure~\ref{noc_topology_arch}(c) shows an example HiMA-NoC for 5$\times$5 tiles. It is made by adding diagonal connections in a mesh NoC. The worst-case inter-tile transfer distance is kept to 4 hops in the 5$\times$5 example. Compared to a fixed NoC that improves traffic conditions for some primitives but worsens for others, HiMA-NoC can be configured in run time to efficiently support different traffic patterns through multi-mode routers (Section~\ref{hima_noc_router}). Using the same simulation setup, HiMA-NoC provides a more scalable speedup than the fixed H-tree, mesh or star NoC, as shown in Figure~\ref{noc_topology_arch}(d). Note that HiMA-NoC does not reduce the amount of traffic, but enhances the tile-to-tile communication latency.

%A notable gap still exists between scalability of HiMA-NoC running DNC and the ideal scaling.
%To further improve HiMA's scalability, we explore algorithmic approaches in Sections~\ref{dist_hist_mem_access} to explore more parallelism opportunities for a fundamental improvement in speedup scalability.
%HiMA's tile-to-tile interconnects are instantiated only between neighboring tiles and the worst-case fan-in is kept constant at 8 regardless of tile count, allowing HiMA to achieve area scalability close to ideal. Detailed area and power of HiMA prototypes are discussed in Section~\ref{area_and_power}. 

\subsection{Submatrix-wise Memory Partition}
\label{mem_optimization}

\begin{figure*}
\centering
\includegraphics[width=0.98\linewidth]{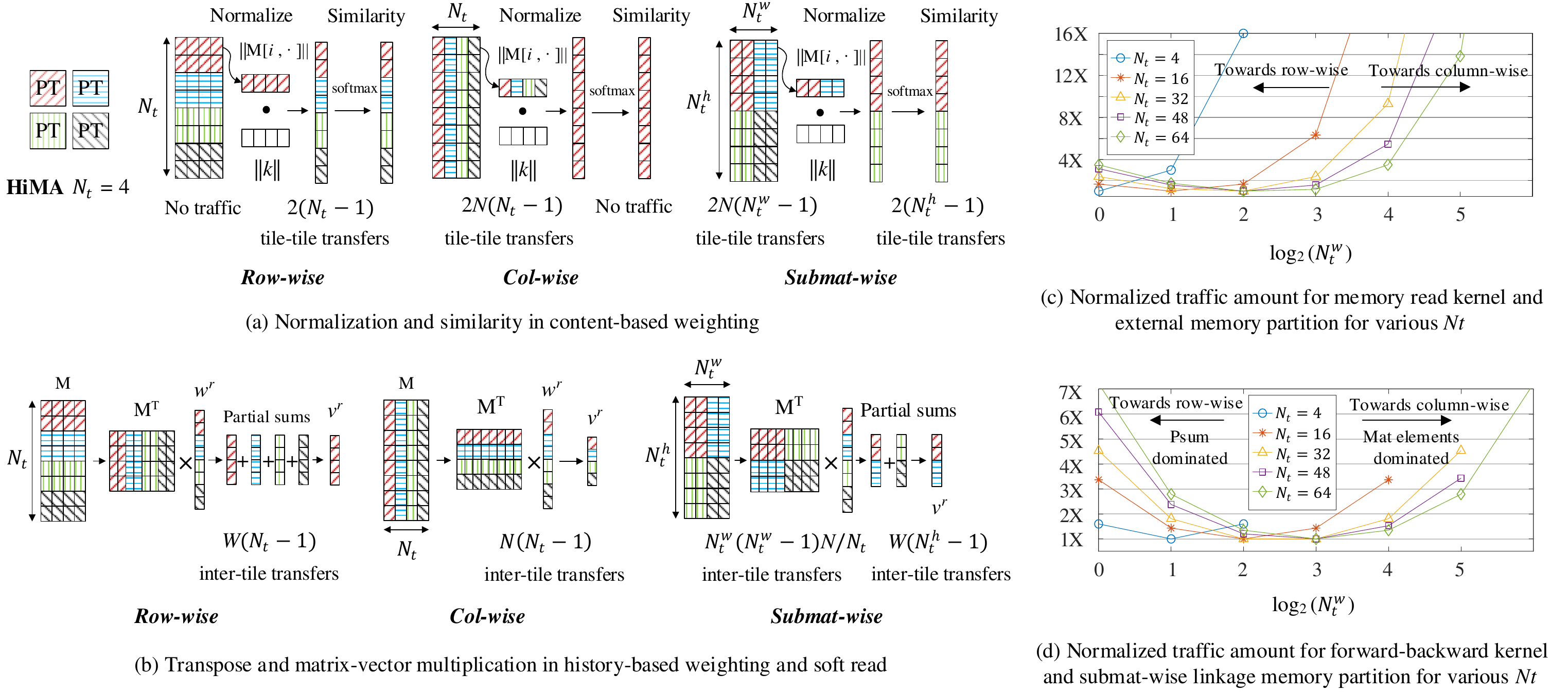}
\caption{External memory partition in a 2$\times$2 tile for (a) content-based weighting, and (b) transpose and matrix-vector multiplication in history-based weighting and soft read; (c) inter-tile traffic for memory read kernel with various external memory partition; (d) inter-tile traffic for forward-backward kernel with various linkage memory partition.}
\label{dnc_mem_arch}
\end{figure*}

\begin{figure}
\centering
\includegraphics[width=0.92\linewidth]{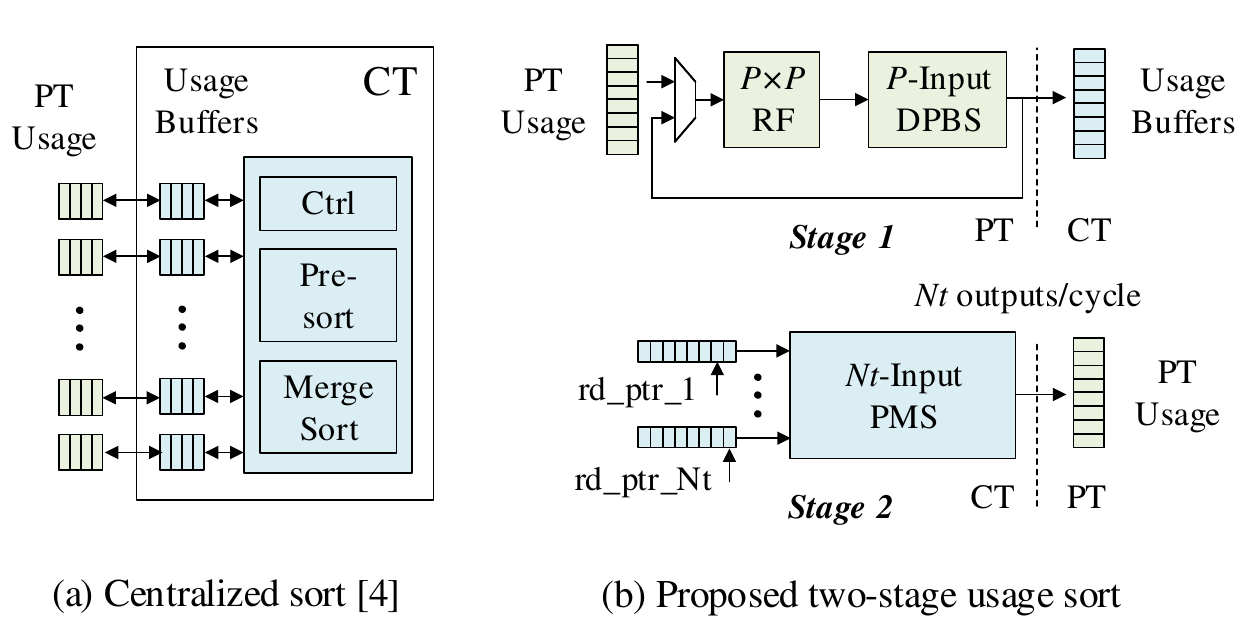}
\caption{Two-stage usage sort.}
\label{two_stage_usage_sort}
\end{figure}

DNC's external memory and state memories need to be partitioned and distributed to the tiles. The partition affects the available access bandwidth, the achievable compute parallelism, and the amount and the patterns of inter-tile data communication. \textsc{S\scriptsize IMBA}\mbox{\cite{simba}}, a state-of-the-art distributed NN accelerator, distributes weights to tiles and efficiently supports convolution and FC workloads for DNNs. \textsc{M\scriptsize ANNA} \cite{manna} partitions external memory row-wise so that each PT receives $N/N_t$ rows of the external memory, where $N_t$ is the number of PTs. \textsc{M\scriptsize ANNA} does not support state memories. 
%State memories were not discussed in \cite{manna}.

HiMA's memory partitions are designed for external memory and new state memories that are nonexistent in NNs or other variants of MANNs like NTM. DNC requires access to various memories concurrently, and the traffic patterns are non-uniform depending on the primitives that are running. There need to be more considerations on memory partition to reduce the amount of traffic.

\subsubsection{External Memory Partition}
\label{ext_mem_optimization}

The external memory is accessed by the access kernels, first in computing content-based weighting including normalization and similarity, followed by memory write or read. Figure~\ref{dnc_mem_arch}(a) illustrates three possible external memory partitions and the inter-tile traffic patterns in computing content-based weighting for a small example of $N_t = 4$ tiles.

The row-wise partition of the external memory eliminates inter-tile transfer for normalization because normalization is computed on a row of memory, which are stored in the same PT. When calculating similarity, one PT produces only a partial sum (psum), and it collects the psums from the rest of $N_t-1$ PTs to compute the global sum, followed by scaling and softmax. The softmax result is then distributed to the $N_t-1$ PTs. Hence the number of inter-tile transfers is $2(N_t-1)$. Alternatively, if we follow the column-wise partition of the external memory, normalization requires $2N(N_t-1)$ inter-tile transfers, but similarity can be computed locally. 

The row-wise and column-wise partitions can be viewed as special cases of a generalized submatrix-wise partition where the external memory is divided into $N_{t}^h$ block rows and $N_{t}^w$ block columns, where $N_t = N_{t}^h\times N_{t}^w$.
%and $N_{t}^h,N_{t}^w \in Z^{+}$
As shown in Figure~\ref{dnc_mem_arch}(a), using submatrix-wise partition, normalization and similarity calculations cost $2N(N_{t}^w-1)$ and $2(N_{t}^h-1)$ inter-tile transfers, respectively.
Based on Eq.~\eqref{extmem_opt1} and given $N \gg N_t$, to minimize the inter-tile traffic, $N_t^w=1$ and $N_t^h=N_t$. In other words, \emph{the row-wise partition of the external memory costs the minimum inter-tile traffic in computing content-based weighting}.

\begin{align}
\label{extmem_opt1}
\operatorname*{argmin}_{N_{t}^h,N_{t}^w} \bigg(2N(N_{t}^w-1) + 2(N_{t}^h-1)\bigg)
\end{align}

The memory write to the external memory requires element-wise operations that can be executed locally by PTs in parallel. The memory read from the external memory requires inter-tile traffic to support matrix transpose and matrix-vector multiplication. Figure~\ref{dnc_mem_arch}(b) illustrates the inter-tile transfer patterns. Similarly, the optimal partition for matrix transpose and matrix-vector multiplication can be derived as Eq.~\eqref{extmem_opt2}.

\begin{align}
\label{extmem_opt2}
\operatorname*{argmin}_{N_{t}^h,N_{t}^w} \bigg(\frac{N_{t}^w(N_{t}^w-1)N}{N_t}+W(N_{t}^h-1)\bigg)
\end{align}

Figure~\ref{dnc_mem_arch}(c) illustrates the external memory partition choices and the impact on inter-tile traffic of memory read kernel for a $N\times W = 1024\times 64$ used in running the bAbI dataset. The five sets of curves correspond to different number of tiles $N_t$, and they cover a range of $N_{t}^w$ choices. Due to the quadratic dependence on $N_{t}^w$, $N_{t}^w$ should generally be kept low to reduce the inter-tile transfers. Therefore, \emph{the row-wise partition of the external memory is advantageous for minimizing the inter-tile traffic in memory read}.

\subsubsection{State Memory Partition}
\label{linkage_mem_optimization}

The state kernels require a set of state memories: usage, linkage, precedence, write weight and read weight. State memories of size $N$ (usage, precedence, write weight) or $N\times R$ (read weight) can be straightforwardly partitioned to $N/N_t$ or $N/N_t \times R$ parts and distributed to the $N_t$ PTs. 

The linkage memory, on the other hand, has a size of $N\times N$. The linkage memory is used by the forward-backward kernel in computing matrix transpose and matrix-vector multiplication. The inter-tile transfer patterns look similar to Figure~\ref{dnc_mem_arch}(b), except that the input matrix is $N\times N$ instead of $N\times W$. Similarly, we find the number of inter-tile transfers based on the generalized submatrix-wise partition and formulate the optimization in Eq.~{\eqref{linkage_opt}}.

\begin{align}
\label{linkage_opt}
\operatorname*{argmin}_{N_{t}^h,N_{t}^w} \bigg(\underbrace{\frac{N_{t}^h(N_{t}^h-1)}{N_t}+N_{t}^w}_\text{Forward} +  \underbrace{\frac{N_{t}^w(N_{t}^w-1)}{N_t}+N_{t}^h}_\text{Backward}\bigg)
\end{align}

The partition choices and the impact on inter-tile traffic of forward-backward kernel are plotted in Figure~\ref{dnc_mem_arch}(d) for a $N\times W = 1024\times 64$ external memory. The results show that both the low-end of $N_{t}^w$ (corresponds to row-wise partition, or transfer psums) and the high-end of $N_{t}^w$ (corresponds to column-wise partition, or transfer matrix elements) are suboptimal. The minimum inter-tile traffic is in between. As an example, for $N_t = 16$, the optimal submatrix partition for the linkage memory is $N_t^h \times N_t^w = 4 \times 4$.

\subsection{Two-Stage Usage Sort}
\label{compute_optimization}

Usage vector sort is a bottleneck primitive. In HiMA, the usage vector is distributed and stored in parts on the PTs. A conventional solution using centralized merge sort \cite{dnc_farm}, as shown in Figure~\ref{two_stage_usage_sort}(a), takes $N\text{log}N$ cycles for a length-$N$ usage vector. To achieve a lower latency, we propose a local-global two-stage sort for the distributed tile architecture: 1) a local usage vector of size $n=N/N_t$ is first sorted by each PT, 2) global merge sort by CT to combine $N_t$ sorted local usage vectors.

We illustrate the two-stage sort for an example of $N_t=4$ tiles and an external memory of $N=1024$ rows. Each PT keeps a local usage vector of length $n=256$. In stage 1 in each PT, a local usage vector is reshaped into a $P\times P$ matrix where $P = \ceil{\sqrt{n}} = 16$. We apply a fast multi-dimensional sorting algorithm (MDSA) \cite{multi_dim_sort} to complete the local sort in only 6 phases. The 2D MDSA sorter is illustrated in Figure~\ref{two_stage_usage_sort}(b), which is composed of a $P\times P$ register file (RF) and a $P$-input dual-mode pipelined bitonic sorter (DPBS) \cite{multi_dim_sort} supporting both ascending and descending order. The 16-input DPBS can be pipelined into $D_{DPBS} = 5$ stages. A length $n = 256$ local usage vector can be sorted in only $6\times (P+D_{DPBS}) = 126$ cycles.

In stage 2, sorted local usage vectors are sent from PTs to CT for global merge sort. CT utilizes $N_t$ memory banks to store the local usage vectors. We apply an $N_t$-input parallel merge sorter (PMS) \cite{parallel_merge_sort} in CT to support $N_t$ outputs per cycle, which are to be written back to the corresponding PTs as shown in Figure~\ref{two_stage_usage_sort}(b). The pointers are updated to keep track of the status of each memory bank. The 4-input PMS can be pipelined into $D_{PMS} = 7$ stages. The global merge sort for the example $N_t = 4$ takes only $n+D_{PMS} = 263$ cycles. With the proposed local-global two-stage sort, usage sort computation latency is reduced to only $6\times (P+D_{DPBS}) + n+D_{PMS} = 389$ cycles compared to $N\text{log}N$ cycles for the centralized merge sort. 

\section{Algorithmic Techniques}

A distributed, tiled architecture provides more opportunities for parallel processing. However, only a subset of DNC primitives like element-wise operations can take the full advantage of distributed processing, while most of the primitives need to operate on the entire external memory or the entire state memories, resulting in excessive traffic and limited hardware scalability. We aim to distribute most of the processing to individual PTs by presenting a distributed version of the DNC model named DNC-D, while minimizing the accuracy loss over DNC.

\begin{figure}
\centering
\includegraphics[width=0.99\linewidth]{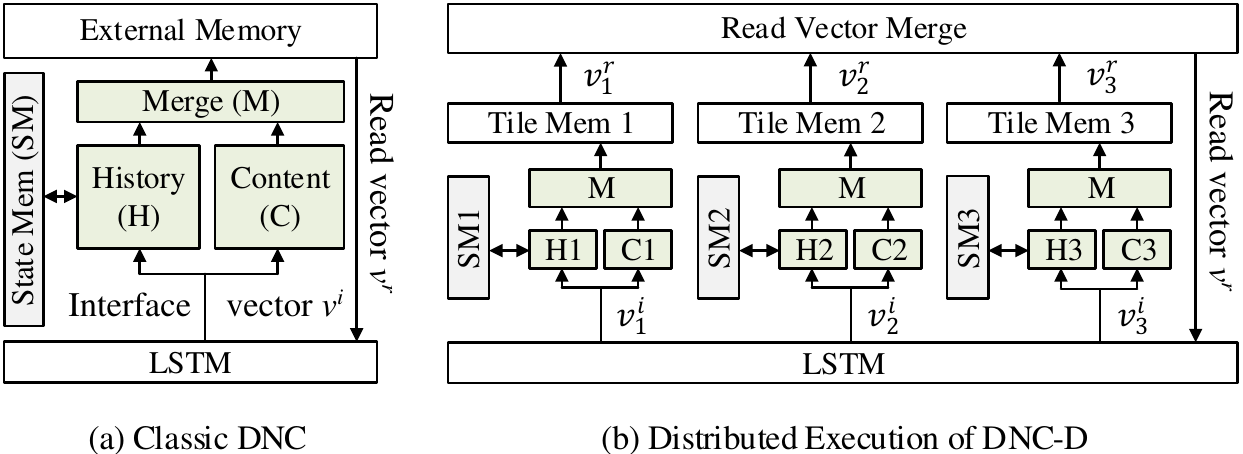}
\caption{Illustration of memory operation in DNC and distributed execution in DNC-D.}
\label{dist_hist_dnc}
\end{figure}

\subsection{Distributed Execution for DNC}
\label{dist_hist_mem_access}

In DNC, the LSTM provides an interface vector to the memory unit as the input and receives a read vector as the output. As shown in Figure~{\ref{dist_hist_dnc}}, in DNC-D the LSTM provides a sub interface vector to each distributed PT instead of broadcasting one global interface vector to all the PTs. Soft read and soft write are executed locally on each PT's local portion of the external memory and state memories.

DNC-D could degrade the inference accuracy. To minimize the loss, we introduce a weighted sum to merge the $N_t$ output read vectors $v_{i}^r$ from the PTs, where $i\in \{1,...,N_t\}$, and compute the final read vector $v^r$ as output to the LSTM as \eqref{rd_vec_merge} below:

\begin{align}
\label{rd_vec_merge}
v^r = \sum_{i=1}^{N_t} \alpha_{i}^rv_{i}^r,
\end{align}

\noindent where the trainable weights $\alpha_{i}^r \in [0,1]$ are determined by the LSTM. The distributed execution offers several advantages: 1) it eliminates the inter-PT communication, 2) it reduces the computations on PTs related to inter-PT data, and 3) it removes the global sort. Without any inter-PT traffic, HiMA achieves nearly optimal speedup scaling as shown in Figure~{\ref{noc_topology_arch}}(d). In Section~{\ref{experiments}}, we study the improved hardware efficiency and the accuracy loss of DNC-D.

\subsection{Approximation Techniques}
We introduce two optional approximation techniques to further reduce the compute complexity. 

\textbf{Usage Skimming:} The usage vectors are collected in computing the write allocation. In practice, we observe that the least significant usage entries have little effect on computation of the write allocation. We propose usage skimming to discard the $K$ smallest usage entries. Usage skimming reduces the complexity of usage sort and write allocation proportionally. In Section~{\ref{experiments}}, we evaluate the inference accuracy impact of usage skimming and show the hardware efficiency improvements.

\textbf{Softmax Approximation:} Softmax is a timing-critical compute block. State-of-the-art softmax approximations include look-up-table (LUT) based \cite{softmax_lut} or piece-wise linear approximation (PLA) based \cite{softmax_piecewise_lut} exponential function. The drawback of the LUT-based is that the number of entries in the table increases exponentially with the input bit width. We combine PLA and LUT approaches: we apply the PLA-based approximation with a small number of line pieces, each of which is an affine function with a slope; and we utilize a LUT of affine functions to store the corresponding function parameters. The design costs only 1 multiply and 1 add.

\section{HiMA Prototype Implementation}

\begin{figure*}
\centering
\includegraphics[width=0.87\linewidth]{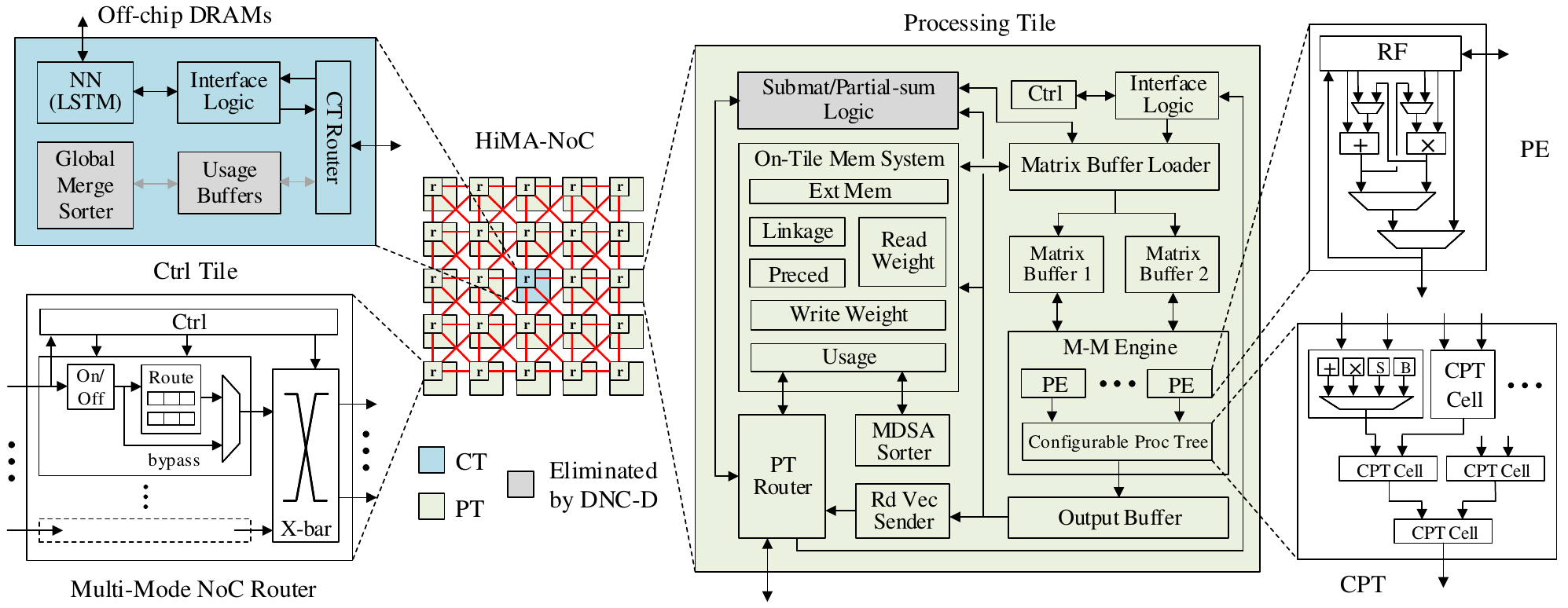}
\caption{HiMA architecture and CT/PT designs.}
\label{dnc_noc_top}
\end{figure*}

Putting everything together, the HiMA architecture is depicted in Figure~\ref{dnc_noc_top}. It is composed of a CT with surrounding PTs connected via HiMA-NoC. HiMA incorporates all the architectural features and the optional algorithmic features.

\textbf{Controller Tile}: CT contains the LSTM and it also executes kernels that require global-level processing. An LSTM implementation employed by \cite{manna} is used in this work and it handles the LSTM inference and communication with off-chip memories. The CT design is illustrated in Figure~\ref{dnc_noc_top}. It sends the interface vectors to the PTs and collects the read vectors from the PTs through routers. The global usage buffers and merge sorter are employed for the 2nd-stage usage sort. Note that using DNC-D, the distributed DNC model, the 2nd-stage usage sort can be eliminated for smaller area.

\textbf{Processing Tile}: A PT's memory system consists of an external memory bank and state memory banks for linkage, precedence, usage, read weighting and writing weighting. The memory partitions are determined based on the submatrix-wise partitions. PT's compute modules support vector and matrix operations for the primitives outlined in Table~\ref{tbl:dnc_kernel_analysis}. Two matrix buffers hold the data for processing from the on-PT memories, the PT router or the interface collector. A matrix buffer loader is used to format and store the data to the corresponding buffers. A matrix-matrix engine (M-M engine) is developed to perform matrix and vector operations. The M-M engine is made of an array of processing elements (PEs) with a configurable processing tree (CPT) to support different sizes of vectors and matrices.

Each PE consists of a small RF to hold the intermediate values. The PE design supports bypass, add, multiply, multiply-then-add or add-then-multiply modes. The CPT consists of multiple stages of compute cells (CPT cells) including adders, multipliers, special function units (SFUs) and bypass routes. It follows a binary tree for reduction and enables faster accumulation. PT also includes a length-$N/N_t$ MDSA sorter for on-tile usage sorting. The proposed architecture provides parallelism through the PE array and the multi-entry RF inside a PE. The size of PE array, the depth of RF inside PE, and the number of CPT stages can be scaled up to provide a higher degree of parallelism.  

\textbf{Multi-Mode NoC Router}:
\label{hima_noc_router}
{Figure~\ref{dnc_noc_top}} illustrates the 8-way multi-mode router that supports different HiMA-NoC modes specified in Figure~\ref{noc_topology_arch}. In addition to the conventional router logic and buffers, input/output ports are controlled by on/off switches to enable traffic only in certain directions. For example, only the east/west ports are enabled for the inner tiles in the ring mode; and only the northeast/southwest ports are enabled in the diagonal mode. Feed-through single-cycle transfer is enabled when the input buffer in the forward direction is empty, bypassing router logic and reducing the latency for non-congested tiles. The multi-mode router is implemented by route LUTs specifically designed to support the proposed modes and a controller that monitors the buffer conditions and generates control signals for each mode.

\section{Evaluations and Benchmarking}
\label{experiments}

%\subsection{Experimental Setup}

We developed a parameterized RTL simulator for HiMA to evaluate its silicon area, inference speed and power consumption. All designs utilize a 32-bit precision for a fair comparison with state-of-the-art MANN accelerators \cite{dnc_farm,manna}. We verified the designs against a functional model of DNC in Python at kernel level as well as system level. To estimate area, we synthesized designs at a 500 MHz clock frequency in a 40nm CMOS technology.
%To estimate speed, inference run time is measured when running DNC and DNC-D.
We used Ansys PowerArtist to obtain power measurements of executing DNC kernels based on switching activities.

\begin{figure}
\centering
\includegraphics[width=.99\linewidth]{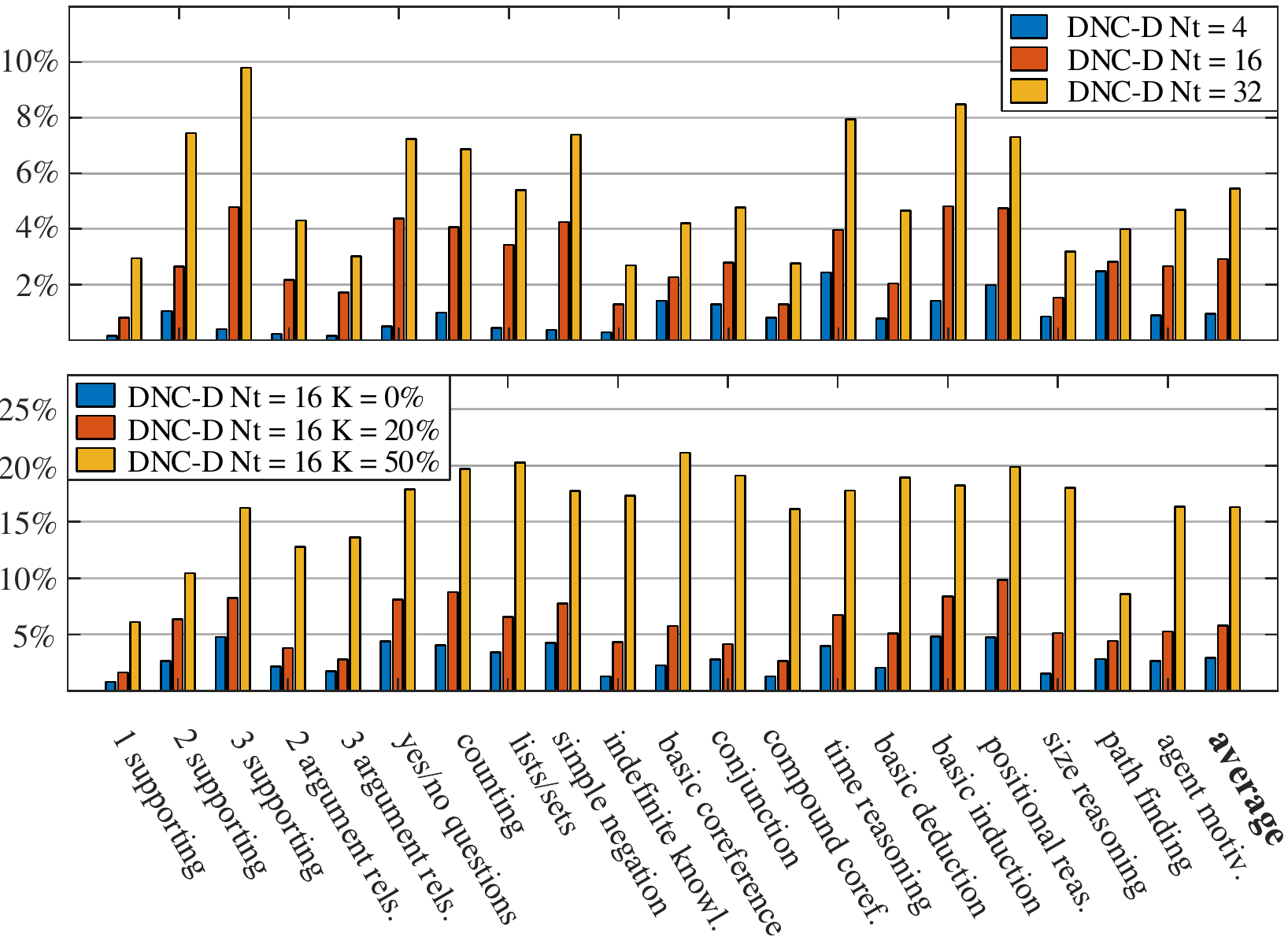}
\caption{DNC-D inference error over DNC for 20 tasks of the bAbI dataset.}
\label{inference_per}
\end{figure}

The HiMA-baseline architecture employs the H-tree NoC used in \cite{manna}. Proposed architectural features, including HiMA NoC, the optimized submatrix-wise memory partition and the two-stage usage sort, are incorporated in the optimized HiMA architectures. HiMA can be further enhanced by the DNC-D model, the usage skimming and the softmax approximation. Based on the architectural and algorithmic features, we create two HiMA architectural prototypes, HiMA-DNC that runs DNC and HiMA-DNC-D that runs DNC-D. Each prototype is equipped with $N_t = 16$ PTs and 1 CT, and supports an external memory of size up to $N\times W = 1024\times 64$ for processing the bAbI dataset. 

%{Each PT has a 64-PE M-M engine and each PE has a depth-64 RF to support submatrix operations of size up to $64\times 64$. For each $Nt$, the PT memory system includes an external memory of 262.144/$N_t$ KBytes, a linkage memory of 4.194/$N_t$ MBytes, and multiple 4096/$N_t$ Bytes memories for the remaining length-$N/N_t$ state memories. }

\begin{figure*}[t]
\centering
\includegraphics[width=0.92\linewidth]{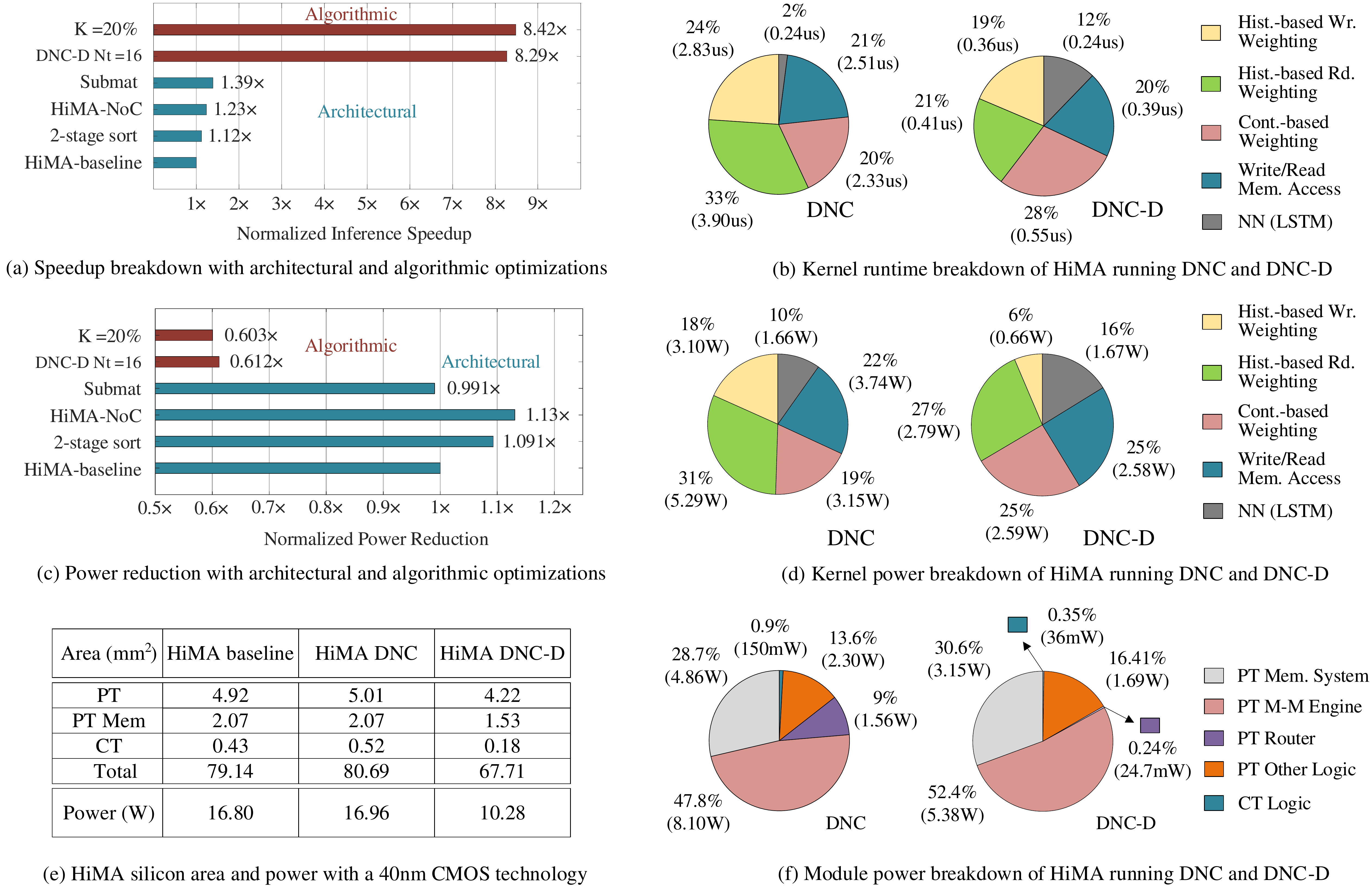}
\caption{HiMA speed, silicon area and power ($N_t = 16$).}
\label{hima_breakdown}
\end{figure*}

\subsection{Inference Accuracy}

To evaluate the inference accuracy of the DNC-D model, we performed simulations using the bAbI dataset and report the error rates over DNC across 20 benchmark tasks in Figure~\ref{inference_per}. The error rate of the DNC-D model increases with the number of distributed tiles $N_t$. If $N_t$ is capped at 32, the average error rate of DNC-D is kept below 6\% over DNC. With a usage skimming rate of $K=$ 20\% and $N_t = 16$, DNC-D demonstrates an error rate of 5.8\% higher than DNC. Further increasing the skimming rate to 50\% increases the error rate above 15\% over DNC. The proposed algorithmic features trade inference accuracy for a higher hardware efficiency. One can select $N_t$ based on the accuracy tolerance.
%e.g., $N_t=4$ gives less than 3\% accuracy drop.
Parameters used in approximations can be selected based on simulations.
%for the application. 

\subsection{Inference Speed}

Figure~\ref{hima_breakdown}(a) itemizes the inference speedup after steps of architectural optimizations over a HiMA-baseline ($N_t = 16$): 1) the two-stage sort provides a 1.12$\times$ speedup over the HiMA-baseline; 2) replacing the H-tree NoC in the HiMA-baseline by the multi-mode HiMA-NoC reduces the communication latency and improves the inference speed to 1.23$\times$ over the baseline. The improvement is mainly due to run time savings of traffic-intensive kernels involving matrix transpose and matrix-vector multiplications, such as linkage, forward-backward and memory read; 3) applying the submatrix-wise partition increases the inference speed to 1.39$\times$ over the baseline, where the speedup is mainly attributed to the reduced traffic amount. These improvements are based on architectural features only. The architecturally optimized HiMA-DNC achieves an inference time of 11.8 $\mu$s per test. Figure~{\ref{hima_breakdown}}(b) shows the kernel run time breakdown in executing DNC. History-based write weighting and read weighting are the most significant, taking 24\% and 33\% of the run time, respectively. 

To further improve the speed, we can apply DNC-D with distributed execution ($N_t=16$). HiMA-DNC-D achieves a 8.3$\times$ inference speedup over the baseline as shown in Figure~\ref{hima_breakdown}(a). The improvement is due to several factors: 1) the elimination of all inter-PT traffic, 2) the computation reduction on PTs, and 3) the elimination of global usage sort. Applying a $K=$ 20\% usage skimming and the softmax approximation increases the inference speedup to 8.4$\times$ over the baseline. The architecturally and algorithmically optimized HiMA-DNC-D with $K=$ 20\% usage skimming shortens the inference time to 1.95~$\mu$s per test. As shown in Figure~{\ref{hima_breakdown}}(b), the run time for history-based write weighting and read weighting in DNC-D are reduced by 87\% and 89\% compared to the run time in DNC, respectively.

\subsection{Silicon Area and Power}
\label{area_and_power}

\begin{figure*}
\centering
\includegraphics[width=0.98\linewidth]{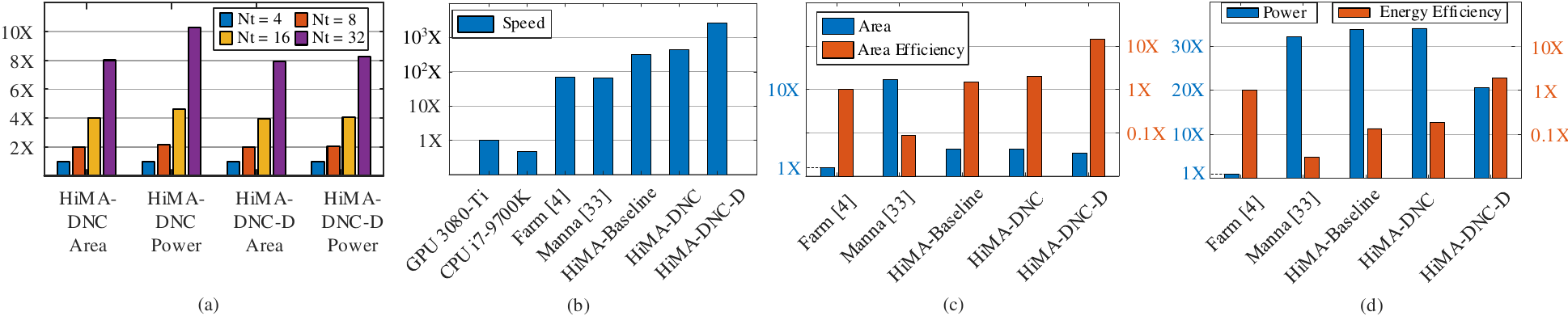}
\caption{(a) Area and power scalability of HiMA-DNC and HiMA-DNC-D to support more tiles and larger external memory; (b)-(d) speed, area and power comparison of HiMA ($N_t = 16$) with state-of-the-art MANN accelerators and GPU/CPU (area efficiency is measured by throughput/area, and energy efficiency is measured by throughput/power).}
\label{speedup_comp}
\end{figure*}

%
%\begin{table}
%	\centering
%	\caption{Silicon area for HiMA ($N_t = 16$) using a 40nm CMOS technology\\}
%	\renewcommand{\arraystretch}{1.2}
%	\begin{tabular}{|c|c|c|c|}
%        \hline
%		\multirowcell{2}{Area (mm$^2$)} & HiMA & HiMA & HiMA \\
%		& baseline & DNC & DNC-D \\ \hline \hline
%		PT & 4.92 & 5.01 & 4.22 \\ \hline
%		PT Memory & 2.07 & 2.07 & 1.53 \\ \hline
%		CT & 0.43 & 0.52 & 0.18 \\ \hline
%		Total & 79.14 & 80.69 & 67.71 \\ \hline \hline
%		Power(W) & 16.80 & 16.96 & 10.28\\ \hline
%	\end{tabular}
%	\label{tbl:hima_area}
%\end{table}

Both HiMA-DNC and HiMA-DNC-D prototypes contain $N_t = 16$ PTs, and they implement all the architectural features. Additionally, HiMA-DNC-D employs a simpler PT and CT due to the elimination of the inter-PT communication and the associated global processing. Figure~\ref{hima_breakdown}(e) compares the silicon area and power consumption of HiMA-DNC and HiMA-DNC-D to HiMA-baseline. HiMA-DNC has a PT area of 5.01~mm\textsuperscript{2}. The architectural features cost an overhead of 1.8\% for the PT over the baseline PT. PT's memory system occupies 2.07~mm\textsuperscript{2}, including an external memory of 16.4~KB, a linkage memory of 262~KB and multiple 256~B state memories.
%for length-$N$ state memories.
The linkage memory and the external memory account for 81.3\% and 4.8\% of the PT memory area, respectively. 

Figure~{\ref{hima_breakdown}}(c) itemizes the power impact of architectural features: 1) the two-stage sort adds 9\% power over the baseline due to the introduction of local sorters in each PT; 2) adopting the multi-mode HiMA-NoC increases the power by another 4\%; 3) applying the submatrix-wise partition reduces the total power to 0.9\% below the baseline, where the power saving comes from the reduced data movement. In all, HiMA-DNC consumes 16.96~W for running a complete DNC inference.

HiMA-DNC-D has a smaller on-PT linkage memory and the centralized sorter is eliminated in CT. It results in a reduced PT area of 4.22~mm\textsuperscript{2} and a reduced CT area of 0.18~mm\textsuperscript{2}. HiMA-DNC-D employs a simpler router that only supports CT-PT traffic as DNC-D eliminates all inter-PT traffic. HiMA-DNC-D uses 16.1\% less silicon area and consumes 39.4\% less power than HiMA-DNC.

Figure~{\ref{hima_breakdown}}(d) and Figure~{\ref{hima_breakdown}}(f) show the kernel and module power breakdown. Notably, DNC-D reduces the power of history-based write weighting by 79\% due to the elimination of global usage sort in CT and the usage transfers between CT and PTs. DNC-D also cuts 98.4\% of the router power because of the elimination of all inter-PT traffic. Since DNC-D allows PT to compute based only on local memories, the computation and traffic reduction result in power savings across all relevant kernels and modules. 

HiMA can be scaled up with more tiles to support a larger external memory and a higher degree of parallelism.
%The inter-tile wiring is limited to neighboring tiles and the increasing CT area for the top-level processing has a limited impact on the total area.
%allowing HiMA to achieve close to the ideal (linear) scaling for both DNC and DNC-D.
As shown in Figure~{\ref{speedup_comp}(a)}, the power of HiMA-DNC grows super-linearly with $N_t$ mainly because of the increased traffic and the related computations on each PT, while DNC-D improves the power scalability close to the ideal (linear) scaling.

\subsection{Comparison with State-of-the-Art Designs}

Figure~\ref{speedup_comp}(b) compares HiMA's performance to the state-of-the-art MANN accelerators as well as an Nvidia 3080-Ti GPU and an Intel Core i7-9700K CPU. The speedup is normalized to the GPU. Figure~\ref{speedup_comp}(c) and Figure~\ref{speedup_comp}(d) compare HiMA's area and power to the MANN accelerators. GPU and CPU are omitted in area and power comparisons since it would be unfair to compare area and power of an accelerator to general-purpose computing platforms. The area and power are normalized to Farm \mbox{\cite{dnc_farm}}. The area is also normalized based on each design's process technology.
%for a fair comparison. 

Farm achieves a 68.5$\times$ faster speed over the GPU. Farm's faster speed is mainly attributed to its small memory size (up to $N=256$) and mixed-signal designs. However, Farm's centralized-memory architecture is not scalable to a larger size to support practical problems and the mixed-signal computation is not yet feasible at a large enough scale. The 16-tile NTM accelerator \textsc{M\scriptsize ANNA} \cite{manna} utilizes an H-tree NoC. It achieves a similar speedup as Farm, but it costs 11$\times$ area and 32$\times$ power to support 20$\times$ larger external memory than Farm. \textsc{M\scriptsize ANNA} still cannot run DNC due to the lack of support for history-based memory access.

HiMA-baseline uses the same H-tree NoC as \textsc{M\scriptsize ANNA} and it supports DNC's history-based memory access. It has a 4$\times$ larger external memory than Farm while using only 3.16$\times$ the area of Farm.
%HiMA-baseline employs 1/4 the external memory as \textsc{M\scriptsize ANNA}.
%Compared to \textsc{M\scriptsize ANNA}, HiMA-baseline is faster and smaller in area because \textsc{M\scriptsize ANNA} uses larger on-tile memories and more compute units to support a 4$\times$ larger external memory than HiMA-baseline.
HiMA-baseline consumes a higher power than \textsc{M\scriptsize ANNA} to support DNC's history-based mechanisms. HiMA-DNC achieves a 1.39$\times$ faster speed over HiMA-baseline thanks to the architectural features. The overhead of the architectural features is almost negligible, which explains why HiMA-DNC uses similar area and power as HiMA-baseline. HiMA-DNC-D takes advantage of the DNC-D model to increase the speed by 8.4$\times$ over HiMA-baseline and reduces the area by 14.4\% and power by 38.8\% over HiMA-baseline.
%The optimized HiMA-DNC-D achieves 38.6$\times$ faster speed, 14.6$\times$ better area efficiency (throughput/area) and 1.87$\times$ better power efficiency (throughput/power), over Farm, the centralized-memory DNC accelerator.
Compared to \textsc{M\scriptsize ANNA} that was designed in a 15nm technology, the 40nm HiMA-DNC-D demonstrates 39.1$\times$ faster speed, 164.3$\times$ better area efficiency and 61.2$\times$ better energy efficiency.

\section{Conclusion}

We present HiMA, a distributed, tile-based accelerator, to efficiently speed up history-based memory access for advanced MANN models like DNC. A multi-mode NoC is designed to support different traffic patterns and improve latency and scalability. Submatrix-wise memory partition is developed to minimize the amount of data movements. To achieve better hardware efficiency, we leverage the tiled architecture to design a two-stage usage sort. To fundamentally improve the efficiency of HiMA's distributed architecture, we distribute not only memory, but also memory operations to the tiles in the form of a new DNC-D model. The HiMA compute kernels can be further optimized by skimming insignificant usage entries and applying an efficient approximation to the softmax function.

We create two HiMA architectural prototypes: HiMA-DNC that runs DNC and HiMA-DNC-D that runs DNC-D. The results show that HiMA-DNC and HiMA-DNC-D achieve 6.47$\times$ and 39.1$\times$ higher speed, 22.8$\times$ and 164.3$\times$ better area efficiency, and 6.1$\times$ and 61.2$\times$ better energy efficiency than \textsc{M\scriptsize ANNA}, the state-of-the-art tiled MANN accelerator for NTM. Compared to an Nvidia 3080Ti GPU, HiMA-DNC and HiMA-DNC-D outperform by up to 437$\times$ and 2,646$\times$ in speed, respectively.

\begin{acks}
This work was supported in part by NSF CCF-1900675.
\end{acks}

%%%%%%% -- PAPER CONTENT ENDS -- %%%%%%%%

\bibliographystyle{ACM-Reference-Format}
\bibliography{sample-base}

\end{document}